\documentclass[12pt]{article} 
\usepackage{amsmath,epsfig}
\usepackage{amsfonts,amssymb}
\usepackage{graphicx}
\usepackage[]{hyperref}
\newtheorem{thm}{Theorem}[section]

\newtheorem{rem}{Remarks}[section]
\newtheorem{claim}{Claim}[section]
\numberwithin{equation}{section}
\newcommand{\Z}{{\mathbb Z}}

\begin{document}

\title {Further discussion of Tomboulis' approach to the confinement 
        problem}
\author { K. R. Ito
          \thanks{E-mail : ito@mpg.setsunan.ac.jp} \\
           Department of Mathematics and Physics\\
           Setsunan University\\
          Neyagawa, Osaka 572-8508, Japan\\
          E. Seiler
          \thanks{E-mail : ehs@mppmu.mpg.de} \\
          Max-Planck Institut f\"ur Physik (Werner-Heisenberg-Institut)\\
          F\"ohringer Ring 6\\
          D-80805 M\"unchen, Germany\\
          }
\date {March 20, 2008}

\maketitle

\begin{abstract}\noindent
      We discuss in some detail certain gaps and open problems in the 
      recent paper by E.~T.~Tomboulis, which claims to give a rigorous proof 
      of quark confinement in 4D lattice Yang-Mills theory for all values of 
      the bare coupling. We also discuss what would be needed to fill the 
      gaps in his proof.\\

\end{abstract}

\section {Introduction}

A paper by E.~T.~Tomboulis that was posted last year on the arXiv 
\cite{tomb} claims to present a rigorous proof of quark confinement 
\cite{wil} in 4D lattice gauge theory. This problem has been fascinating 
many physicists since the last century and still remains an important open 
problem; in fact, its resolution probably would represent a first step 
towards the solution of one of the millennium problems posed by the Clay 
Mathematics Institute \cite{clay}.  In a comment posted on the arXiv 
\cite{ito_ehs} we pointed out that in our opinion the proof in 
\cite{tomb} depended on a claim which remains to be proved. The purpose of 
this public comment was to prevent confusion in the community about the 
question whether confinement on the lattice was proven or not. Tomboulis 
then posted a reply \cite{tomb2} on the arXiv stating that the presumed 
gap found by us was simply due to a misunderstanding of his argument. 

We appreciate that Tomboulis gave a concise and clear exposition of the 
logic of his argument in his reply. But we still think that our original 
criticism stands and in this note we try to give a more detailed 
explanation of the main open problems we see in Tomboulis' approach. We 
also discuss possible ways to complete his proof.

\section {Tomboulis' claimed theorem}

To make this note self-contained, we summarize Tomboulis' notation and 
his arguments with some simplifications:

\begin{enumerate}
\item  $\Lambda \subset \Z^{4}$ is a box in $\Z^{4}$ of size 
       $L_{1}\times \ldots \times L_{4}$ with center at the origin, 
       where $L_{i}=ab^{n_{i}}$, $n_{i}>>1$, $i=1,\ldots,4$ and 
       $a,b$ are positive integers larger than 1.
\item  $\Lambda^{(k)} \subset \Z^{4}$ is the box in $\Z^{4}$ 
       of each side length $L_{i} b^{-k}$ obtained from $\Lambda$ 
       by $k$ steps of the renormalization transformation of 
       Migdal-Kadanoff type ($\Lambda=\Lambda^{(0)}$). 
       Periodic boundary conditions are employed in all 
       directions of $\mu=1,\ldots, 4$. (actually periodic boundary 
       conditions in the directions of $\mu=3,4$ would be enough.)
\item  $f(\{c_{j}\},  U)=1+\sum_{j\neq 0} c_{j}d_{j}\chi_{j}(U)$ 
        where $d_{j}$ is the dimension of the representation $\chi_{j}$, 
        and we assume $U\in G=SU(2)$. Moreover 
        $U=U_{p}=\prod_{b\in\partial p} U_{b}$ for a plaquette 
        $p\subset \Lambda$ which consists of bonds $b\in \partial p$ 
        oriented positively with respect to $p$.
\end{enumerate}

We start with
\begin{subequations}
\begin{eqnarray}
\exp\left [ \frac{\beta}{2}\chi_{1/2}(U) \right ]
       &=&F_{0}(0)f(\{c_{j}\}, U) \\
f(\{c_{j}\},U) &=& 1+\sum_{j\neq 0} c_{j}d_{j}\chi_{j}(U) \\
F_{0}(0) &=& \int \exp\left[  \frac{\beta}{2} \chi_{1/2}(U)\right ] dU
\end{eqnarray}
\end{subequations}
where $dU$ is the Haar measure of $G=SU(2)$. We then apply the
renormalization group (RG) recursion formulas of Migdal-Kadanoff (MK) 
type  \cite{mk} with some modifying parameters. If there are no such 
parameters (the standard recursion formula), we have
\begin{eqnarray*}
f^{(n-1)}(U) &=&f(\{c_{j}(n-1)\},U) \\
             &=& 1+\sum_{j\neq 0} c_{j}(n-1)d_{j}\chi_{j}(U)\\
             &\to& f^{(n)}(U)=f(\{c_{j}(n)\},U)
\end{eqnarray*}
where
\begin{subequations}
\begin{eqnarray}
f^{(n)} (U) &=& 
            \frac{1}{F_{0}(n)} \int
              [ \, \underbrace{f^{(n-1)}(UU_{1})
                   f^{(n-1)}(U_{1}^{-1}U_{2})
                   \cdots 
                   f^{(n-1)}(U_{b^{2}})}_{b^{2}}
                \,  ]^{b^{2}}
                   \prod dU_{k}         \nonumber \\
             & &             \label{Convolution}\\
    F_{0}(n) &=& \left(\int [f^{(n-1)}(U)]^{b^{2}} dU\right )^{b^{2}}
\end{eqnarray}
\end{subequations}
or equivalently
\begin{subequations}
\begin{eqnarray}
c_{j}(n) &\equiv & (\hat{c}_{j}(n))^{b^{2}}
         =  \frac{F_{j}(n)}{F_{0}(n)}   \label{def_of_r} \\
    F_{j}(n) &=&
           \left( \int [f^{(n-1)}(U)]^{b^{2}}\frac{\chi_{j}(U)}{d_{j}}
            dU \right)^{b^{2}}
\end{eqnarray}
\end{subequations}
in terms of the coefficients of the character expansions. Then  
\cite{ito}: 
\begin{thm} \label{ito}
     For $D\leq 4$ and for $G=SU(N)$ or $G=U(N)$, 
     $$\lim_{n\to\infty} c_{j}(n)=0\quad\quad {\rm for}\; j\neq 0$$
\end{thm}

These recursions are just approximate and yield upper bounds for the 
partition functions \cite{tomb}. Tomboulis \cite{tomb} then introduces two 
parameters $\alpha\in(0,1]$ and $t>0$ and a function $h(\alpha,t) \in 
(0,1]$, designed so that this transformation becomes numerically exact for 
the partition function:

\begin{subequations}
\begin{eqnarray}
Z &=& \int dU_{\Lambda} \prod_{p\subset \Lambda } f(\{c_{j}\},U_{p})  \\
  &=& [F_{0}(1)^{h(\alpha,t)}]^{|\Lambda^{(1)}|}
      \int dU_{\Lambda^{(1)}} \prod_{p\subset \Lambda^{(1)}} 
      f(\{\tilde{c}_{j}(\alpha)\},U_{p}) \\
  &=& \tilde{Z}_{1} (\tilde{c}(\alpha),t) 
\end{eqnarray}
\end{subequations}
where $dU_{\Lambda}=\prod_{b\in\Lambda} dU_{b}$, 
\begin{subequations}
\begin{eqnarray}
\tilde{Z}_{1} (\tilde{c}(\alpha),t) 
      &=& [F_{0}(1)^{h(\alpha,t)}]^{|\Lambda^{(1)}|} Z_{1}\\
Z_{1} &=& \int dU_{\Lambda^{(1)}} \prod_{p\subset \Lambda^{(1)}} 
           f(\{\tilde{c}_{j}(\alpha)\},U_{p}) 
\end{eqnarray}
\end{subequations}
and as usual, $U_{p}=\prod_{b\in\partial p} U_{b}$
are plaquette actions defined as the product of 
group elements $U_{b}=U_{(x,x+e_{\mu})}\in G$
attached to the (oriented) bonds $b\in \Lambda$. Moreover
\begin{subequations}
\begin{eqnarray}
h(\alpha,t) &=& \exp[-t(1-\alpha)/\alpha] \\
\tilde{c}_{j}(\alpha)
            &=& \tilde{c}_{j}^{(1)}(\alpha)= \alpha c_{j}(1) \\
 c_{j}(1)   &=& \frac{F_{j}(1)}{F_{0}(1)}\,,
\end{eqnarray}
\end{subequations}
(where without loss of generality we have chosen one particular form of 
the function $h$, as suggested in \cite{tomb}). Given $t>0$, $\alpha\in 
[0,1]$ is chosen so that the above relation becomes exact; thus $\alpha$ 
has to be considered as a function of $t$.

The author of \cite{tomb} then introduces a 
vortex sheet $V=\{v\subset \Lambda\}$  \cite{thooft,mack} 
which is a collection of plaquettes 
$\{v=\{x_{0}+n_{3}e_{3}+n_{4}e_{4},x_{0}+e_{1}+n_{3}e_{3}+n_{4}e_{4}, 
x_{0}+e_{2}+n_{3}e_{3}+n_{4}e_{4},
x_{0}+e_{1}+e_{2}+n_{3}e_{3}+n_{4}e_{4}\}, n_{i}=0,1,,\ldots, L_{i},
(i=3,4) \}$, 
i.e. a plaquette $p=(x_{0}, x_{0}+e_{1}, x_{0}+e_{1}+e_{2}, 
x_{0}+e_{2})$ in an $x_{1}-x_{2}$ plane and its translates 
along the axis normal to $p$ (say, 3rd and 4th axis). 
Following \cite{tomb} we define
\begin{eqnarray}
Z^{(-)} &=& \int dU_{\Lambda} \prod_{p\subset \Lambda } 
           f(\{c_{j}\},(-1)^{\nu(p)}U_{p}) 
\end{eqnarray}
where 
\begin{equation}
\nu(p)=\left\{ \begin{array}{ll}
                  0  & \mbox {if $p\notin V$}\\
                  1  &  \mbox {if $p\in V$}\\
          \end{array}
         \right .
\end{equation}
and then 
\begin{eqnarray}
Z^{(-)} &=& \int dU_{\Lambda} \prod_{p\subset \Lambda\backslash V} 
              (1+\sum_{j\neq 0} c_{j}d_{j}\chi_{j}(U_{p}))
                     \nonumber \\
      & & \times      \prod_{q\subset V} 
              (1+\sum_{j\neq 0} (-1)^{2j}c_{j}d_{j}\chi_{j}(U_{q}))
\end{eqnarray}

Note that the position of the vortex sheet $V\subset \Lambda$ can 
be freely moved in the $x_{1}-x_{2}$ plane by gauge 
invariance. 

First of all, note that the coefficients of $\chi_{j}(U)$ in 
$\prod_{p} (1+\sum_{j\neq 0} c_{j}d_{j}\chi_{j}(U_{p}))$
and 
$$
\prod_{p} (1+\sum_{j\neq 0} c_{j}d_{j}\chi_{j}(U_{p}))
                +\prod_{p} (1+\sum_{j\neq 0} c_{j}d_{j}
                        \chi_{j}((-1)^{\nu(p)}U_{p}))
$$
are nonnegative. The main claims in \cite {tomb} are: 
\begin{claim}  \label{MainClaim}
(1) There exist $t$, $t^{+}>0$ and functions $\alpha$ and 
$\alpha^{+}$ such that 
$$
\frac{Z^{(-)}_{\Lambda}(\{c_{j}\})} {Z_{\Lambda}(\{c_{j}\})}
=
\frac{Z^{(-)}_{\Lambda^{(n)}}(\{\tilde{c}^{(n)}_{j}(\alpha^{+}(t^{+}))\})}
   {Z_{\Lambda^{(n)}}(\{\tilde{c}^{(n)}_{j}(\alpha(t))\})}
$$
where
$$
\tilde{c}^{(n)}_{j}(\alpha(t))=\alpha(t) c_{j}(n). 
$$
(2) For large $n$ such that $\{c_{j}(n)\}$ are sufficiently small, there 
exists a $t_*\geq 0$ (the same for numerator and denominator!) such that 
$$
\frac{Z^{(-)}_{\Lambda}(\{c_{j}\})} {Z_{\Lambda}(\{c_{j}\})}
=
\frac{Z^{(-)}_{\Lambda^{(n)}}(\{\tilde{c}^{(n)}_{j}(\alpha(t_*))\})}
       {Z_{\Lambda^{(n)}}(\{\tilde{c}^{(n)}_{j}(\alpha(t_*))\})}
$$
\end{claim}

If this were correct, the following would be true: 
$$
\frac{Z^{(-)}_{\Lambda}(\{c_{j}\})} {Z_{\Lambda}(\{c_{j}\})}
=
\frac{Z^{(-)}_{\Lambda^{(n)}}(\{\tilde{c}^{(n)}_{j}(\alpha(t))\})}
       {Z_{\Lambda^{(n)}}(\{\tilde{c}^{(n)}_{j}(\alpha(t))\})}
$$
Since for the (unmodified) Migdal-Kadanoff RG described
above, $\{c_{j}^{(n)} \geq 0 \}$ tends to the high 
temperature fixed point(i.e. $\{c_{j}(n)\} \to 0$)  as $n\to\infty$
if the dimension is $\leq4$, whether $G$ is abelian or non-abelian
(see the remark below and \cite{ito}), this would mean strict 
positivity of 't\,Hooft's string tension and thus 
establish permanent confinement of quarks in the sense of Wilson
at least for all values of the bare coupling constant \cite{mack,bose} in 
4 dimensional lattice gauge theory, thereby solving part of a longstanding 
problem in modern physics (the missing part, as Tomboulis makes clear, 
would still be the proof that there is a continuum limit in which 
confinement persists).

As pointed out in our earlier note \cite{ito_ehs}, this cannot possibly 
be correct, however, since there exists a deconfining transition in 4D 
lattice gauge theory based on abelian gauge groups. But it is meaningful 
for the future study of the confinement problem to ask what is going wrong 
in reaching this conclusion.

The original idea to study vortex free energies in order to understand 
confinement was formulated by Mack and Petkova in \cite{mack}. They 
fix boundary variables $G_{\partial\Lambda}
    \equiv \{g_{b}\in G ; b=(x,x+e_{\mu})\in\partial \Lambda\}$
and take the maximum of $Z^{(-)}_{\Lambda}/Z_{\Lambda}$ over 
$G_{\partial \Lambda}$.  On the other hand  't\,Hooft's string tension
is defined using (twisted) periodic boundary conditions at $\partial 
\Lambda$, see \cite{bose}. The following are well-known facts:

\begin{thm} Define 
$$F\equiv - \log\frac{Z^{(-)}_{\Lambda}}{Z_{\Lambda}}$$ 
(1) Let $G=U(1)$ or $G=SU(2)$. Then there is a $\beta_1>0$ such that 
for $\beta<\beta_1$ 
$$
\lim_{L_3L_4\to\infty}\frac{1}{L_3L_4}F\le
{\rm const}\,\exp[-\sigma L_{1}L_{2}]
$$
(area decay law). $\sigma>0$ is, if chosen maximally,  is 't\,Hooft's 
string tension. \\
(2) Let $G=U(1)$. Then there is a $\beta_0<\infty$ such that
for
$\beta>\beta_0$   
$$
\lim_{L_3L_4\to\infty}\frac{1}{L_3L_4}F\ge {\rm const}\, \frac{1}{L_1L_2}
$$
(deconfinement).
\end{thm}
(1) is a standard result of the convergent high-temperature expansion 
(see for instance \cite{seiler2}). The analogous statement for $SU(N)$
holds with $Z^{(-)}$ replaced by $Z^\omega$, the partition function 
twisted by an element $\omega$ of the center of $SU(N)$. (2) is a 
variation of the results of \cite{FrSp} and \cite{Guth} (see \cite{ito_ehs2}).

\begin{rem} (1) The introduction of $0<\alpha\leq 1$ 
into $1+\sum_{j\neq 0} c_{j}d_{j}\chi_{j}(U)$ does not 
violate conditions (positivity, analyticity, class functions  etc.) 
on $f^{(n)}(v)$ in \cite{ito} since 
$$1+\sum_{j\neq 0} \alpha c_{j}d_{j}\chi_{j}(U)
  =(1-\alpha)+\alpha\left(1+\sum_{j\neq 0}
		       c_{j}d_{j}\chi_{j}(U)\right)$$
and $0<\alpha\leq 1$. Then $\{c_{j}(n)\geq 0\}$ tends to 0 as 
$n\to\infty$ (if $r=1$, see Section 3.3). \\
(2) The conjecture raised in \cite{mack} is proved rigorously in 
\cite{bose}. Namely 't\,Hooft's string tension is smaller than or equal 
to Wilson's string tension.\\
(3) The center of  $G=U(1)$ is again $U(1)$. Then we consider 
$U(p)=\cos(\theta_{p})$ and $U_{\omega}(p)=\cos(\theta_{p}+\omega)$
 for $p\in V$. $U_{\omega}(p)=-U(p)$ for $\omega=\pi$.\\
(4) The argument in  \cite{FrSp} uses the Fourier transformation of 
$e^{\beta \cos \theta}$ and Poincar\'e's lemma. Since $\Lambda$ is a 
torus, Poincar\'e's lemma has to be replaced by the Hodge decomposition in 
order to adapt the arguments in \cite{FrSp} or \cite{Guth}. See 
\cite{ito_ehs2} and \cite{seiler2}.
\end{rem}

\section {Tomboulis'  Proof revisited}

We now analyze the arguments in \cite{tomb} in more detail. First, using 
the fact that the partition function $Z=Z_{\Lambda}$ increases by the 
Migdal-Kadanoff (MK) recursion formula \cite{tomb}, Tomboulis introduces 
two interpolation parameters $\alpha$ ($Z$ increases as $\alpha\nearrow 
1$) and $t$ (the factor $[F_{0}(n)]^{h(\alpha,t)}$ decreases as $t$ 
increases). Then he claims that there exist functions $\alpha(t)$ and 
$\alpha^{+}(t)$ such that 
\begin{subequations} \begin{eqnarray} 
Z_{\Lambda}(\{c_{j}\})
     &=&[F_{0}(1)]^{h(\alpha(t),t)|\Lambda^{(1)
}|}
     Z_{\Lambda^{(1)}}(\{\tilde{c}_{j}(\alpha(t))\}) \\
Z_{\Lambda}^{+}(\{c_{j}\})
    &=&[F_{0}(1)]^{h(\alpha^{+}(t),t)|\Lambda^{(1)}|}
     Z_{\Lambda^{(1)}}^{+}(\{\tilde{c}_{j}(\alpha^{+}(t))\})
\end{eqnarray}
\end{subequations}
where
$$ Z^{+}=\frac{1}{2}(Z+
Z^{(-)}) $$
and the right hand sides are manifestly independent of $t$. 

The author of \cite{tomb} then claims is that there exists a $t_{*}>0$
such that 
\begin{equation}
 \alpha (t_{*})=\alpha^{+} (t_{*})  
                   \label{XXX}
\end{equation}
which yields
\begin{equation}
\frac{Z_{\Lambda}^{+} (\{c_{j}\})} {Z_{\Lambda}(\{c_{j}\})}
=\frac{Z_{\Lambda^{(1)}}^{+} (\{\tilde{c}_{j}\})}
      {Z_{\Lambda^{(1)}} (\{\tilde{c}_{j}\})}
\end{equation}
where
$$  \tilde{c}_{j}= \tilde{c}_{j}(\alpha(t_{*})).$$

The procedure is then iterated to produce an analogous relation involving 
the $n-$th iteration of the  MK RG transformation. 

Two arguments are given in \cite{tomb} for this proof. 
\begin{description}
\item{large $\beta$ region:} p.24--p.26,
      based on the arguments in pages 14,15,22, 23 and Appendix B.
\item{small $\beta$ region:} p.26--p.27 and Appendix C.
\end{description}

The first argument is used to drive the system to the 
high-temperature  (small $\beta$)  region  by the MK recursion 
formulas, and the second argument depends on the implicit 
function theorem to derive the main claim which yields strict 
positivity of 't\,Hooft's string tension.

\subsection{Outline of the Proof in \cite{tomb} }

To understand the proof in in \cite{tomb}, we streamline 
the arguments in \cite{tomb} by extracting key parts from \cite{tomb}
(some equations are added below by the present authors). But before doing 
so, we have to stress that Tomboulis is modifying the MK recursion by 
introducing a further parameter $r\in(0,1]$ (Eq. (2.19) in \cite{tomb}), 
which is the source of some serious problems, because any choice $r<1$ it 
changes the MK RG flow for large $\beta$ drastically (see below). 

$r$ is introduced by replacing (\ref{def_of_r}) with 
$c_{j}(n)=\hat{c}_{j}(n)^{b^{2}r}$ (this amounts to replacing the 
$b^{2}$ convolutions in Eq.(\ref{Convolution}) by $b^{2}r$ convolutions). 

\begin{quotation} \noindent {\bf Quotation (I):} [page 14] $$ 
\frac{\partial \alpha_{\Lambda,h}^{(m)}(t,r)}{\partial t} 
=v(\alpha_{\Lambda,h}^{(m)}(t,r),t,r)  
\mbox{\hspace*{3.6cm}}(3.25) $$
where 
$$ 
v(\alpha,t,r)\equiv - \frac{\partial h(\alpha,t)/\partial t}
                {\displaystyle 
                   \frac{\partial h(\alpha,t)}{\partial \alpha}
                        + A_{\Lambda^{(m)}}(\alpha,r)}
                         \mbox{\hspace*{2.4cm}} (3.26) 
$$
\end{quotation}
with $\partial h(\alpha,t)/\partial t=-[(1-\alpha)/\alpha]h(\alpha,t)<0$ 
and 
\begin{quotation}\noindent
with
\begin{eqnarray*}
A_{\Lambda^{(m)}}(\alpha,r)
 &=&\frac{1}{|\Lambda^{(m)}|\log F_{0}^{U}(m)}
    \frac{\partial}{\partial \alpha}
    \log Z_{\Lambda^{(m)}}\left(\{\tilde{c}_{j}(m,\alpha,r)\}\right)>0\\
     & & \mbox{\hspace{6.8cm}} (3.27)
\end{eqnarray*}

\noindent
[ page 15] We require that  
$$\delta'<\alpha_{\Lambda,h}(t,r)<1-\delta 
\mbox{\hspace*{5.0cm}}  (3.30) $$
{\it \ldots by choosing $r$ to vary, if necessary, 
away from unity in the domain}
$$1>r>1-\varepsilon \mbox{\hspace*{6.4cm}}   (3.31)$$
where $0<\varepsilon<<1$ with $\varepsilon$ independent of 
$|\Lambda^{(m)}|$.
With (3.30) in place, (3.25) and (3.29) imply (Appendix B) that
$$
\frac{\partial }{\partial t} \alpha_{\Lambda,h}(t,r)
      > \eta_{1}(\delta) >0  \mbox{\hspace*{4.7cm}} (3.32)
$$
[page 23]  {\it by letting the parameter $r$ vary, if necessary}, 
in (3.31). From this, it follows that the analog of (3.32):
$$
\frac{\partial }{\partial t} \alpha_{\Lambda,h}^{+}(t,r)
      > \eta_{1}^{+}(\delta^{+}) >0, \ldots 
\mbox{\hspace*{3.7cm}}   (4.25)
$$
holds \ldots
\end{quotation}
(italics introduced by us for emphasis).

\noindent 
\begin{quotation}
\noindent
{\bf Quotation (II):} [page 25]  $\ldots$ (5.4), taken at general $r$, 
implies that $\ldots$
$$
\left|\alpha^{+}_{\Lambda,h}(t,r)-\alpha_{\Lambda,h}(t,r)\right|
    \leq O\left(\frac{1}{|\Lambda_{1}|}\right)
\mbox{\hspace*{3.0cm}}  (5.6)
$$
$\ldots$. For any given $t_{1}^{+}$, choose $t_{1}$ in 
$\tilde{Z}_{\Lambda,h}(\beta,\alpha_{\Lambda,h}(t_{1}),t_{1})$ 
so that 
  $$h(\alpha_{\Lambda,h}(t_{1}),t_{1})
      =h(\alpha_{\Lambda,h}^{+}(t_{1}^{+}),t_{1}^{+})
    \mbox{\hspace*{3.8cm}}    (5.7)$$
This is clearly always possible by (3.32) and (4.25), and by (5.6).\\

\noindent
[top of page 26]   We may now iterate this procedure 
$(n-1)$ decimation steps $\ldots$,  at each step choosing 
$t_{m}$ and $t^{+}_{m}$ such that 
  $$h(\alpha_{\Lambda,h}^{(m)}(t_{m}),t_{m})
      =h(\alpha_{\Lambda,h}^{+(m)}(t_{m}^{+}),t_{m}^{+})
    \mbox{\hspace*{3.6cm}}      (5.9) $$
Carrying out $\ldots$ one obtains
$$
\left(1+\frac{Z_{\Lambda}^{(-)}}{Z_{\Lambda}} \right )
   =\frac{2\tilde{Z}_{\Lambda^{(n)}}^{+}(\beta,h,
                     \alpha^{+(n)}_{\Lambda,h}(t^{+}),t^{+})}
         {\tilde{Z}_{\Lambda^{(n)}}(\beta,h,
                     \alpha^{(n)}_{\Lambda,h}(t),t)}
\mbox{\hspace*{2.6cm}}  (5.10)
$$
\end{quotation}

\begin{quotation}
\noindent 
{\bf Quotation (III):} [middle of page 26] Next, consider (5.10) rewritten 
as
\begin{eqnarray*}
\left(1+\frac{Z_{\Lambda}^{(-)}}{Z_{\Lambda}} \right )
  &=&\left(\frac{{Z}_{\Lambda^{(n-1)}}^{+}}
            {\tilde{Z}_{\Lambda^{(n-1)}}^{+}
               (\beta,h, \alpha^{(n)}_{\Lambda,h}(t),t)}
    \right )
   \left(1+ \frac{{Z}_{\Lambda^{(n)}}^{(-)}(\{\tilde{c}_{j}(\ldots)\})}
            {\tilde{Z}_{\Lambda^{(n)}}
               (\{\tilde{c}_{j}(\ldots)\})}
   \right) \\
 & & \mbox{\hspace*{7.0cm}} (5.13)
\end{eqnarray*}
$\ldots$. It is then natural to ask whether there exists a value
$t=t_{\Lambda,h}^{(n)}$ such that
$$
  \tilde{Z}_{\Lambda^{(n-1)}}^{+}
               (\beta,h, \alpha^{(n)}_{\Lambda,h}(t),t)= 
{Z}_{\Lambda^{(n-1)}}^{+}
 \mbox{\hspace*{4cm}}  (5.14)
$$
Note that the graphs of $\alpha_{\Lambda,h}^{(n)}(t)$ and 
$\alpha_{\Lambda,h}^{+(n)}(t)$ must intersect at 
$t_{\Lambda,h}^{(n)}$. 
A unique solution to (5.14) indeed exists as shown in Appendix C 
provided 
$$
A_{\Lambda^{(n)}}(\alpha,r)\geq A_{\Lambda^{(n)}}^{+}(\alpha,r)
\mbox{\hspace*{5.4cm}} (5.15)
$$

 \noindent 
\end{quotation}
(The above assertion is to be interpreted as affirmative.)
\begin{quotation}\noindent
 [near top of page 27] 
 {\bf V.1} Consider $n$ successive decimation steps performed 
  according the scheme (4.28). Assume that there 
 is an $n_{0}$ such that the upper bound coefficients $c_{j}^{U}(n)$
 become sufficiently small for $n>n_{0}$. \\
\noindent 
[Appendix C, page 44] With the notation 
$$
\Phi^+_{\Lambda^{n}}(\alpha)\equiv \frac{1} {\ln\, 
F_0^U(n)}\frac{1}{|\Lambda^{(n)}|}\ln Z^+_{\Lambda^{n}}(\{\tilde 
c_j(n,\alpha)\}) \mbox{\hspace*{1.8cm}} {\rm (C.5)}
$$
we now define 
$$
\Psi(\lambda,t)\equiv  h(\alpha(t),t)+
(1-\lambda)\Phi^+_{\Lambda^{n}}(\alpha^+_h(t_I))+\lambda 
\Phi^+_{\Lambda^{n}}(\alpha_h(t))-\Phi^+_{\Lambda^{n-1}}
$$
\hspace*{10.1cm}  {\rm (C.6)}

\noindent
and consider the equation
   $$\Psi(\lambda,t)=0   \mbox{\hspace*{7.7cm}} {\rm (C.7)}$$
At $\lambda=0$,  eq.(C.7) is solved by setting $t=t_{0}$ $\ldots$. 
By the implicit function theorem, if grad $\Psi$ is 
continuous  and $\partial \Psi/\partial t\neq 0$, there exists
a branch $t(\lambda)$ through $(0,t_{0})$ on a $\ldots$. 
Thus, from (C.8), $t(\lambda)$ \ldots  extends to the solution 
$t(1)>t_{0}$.
\end{quotation}

Quotation (III) is  the main result and quotation (II) is used 
as the bridge to reach the main conclusion.
The summary of the argument in \cite{tomb} is:
\begin{description}
\item{(I)} If $r<1-\varepsilon$ then 
         $\alpha<1-\delta<1$ and    $\alpha^{+}<1-\delta<1$.
        ($\delta>0$).    Thus 
$v=\partial\alpha(t,r)/\partial t$ 
and  $v^{+}=\partial\alpha^{+}(t,r)/\partial t$ 
are bounded from below by strictly positive constants
$\eta_{1}>0$ and $\eta_{2}>0$.
\item {(II)} Since $|\alpha^{+}(t^{+})-\alpha(t)|= 
       O(|\Lambda_{1}|^{-1})$,  
       the existence of $t_{1}$ such that 
       $$h(\alpha_{\Lambda,h}(t_{1}),t_{1})
      =h(\alpha_{\Lambda,h}^{+}(t_{1}^{+}),t_{1}^{+})
       $$ 
       follows by a {\it shift } of $t$.
        Moreover it says that  one can find $t=t_{m}$ successively 
        to obtain  
        $Z^{+}_{\Lambda^{(m)}}(\alpha^{+(m)}(t^{+}))
             /Z_{\Lambda^{(m)}}(\alpha^{(m)}(t))$.
        This proves Claim \ref{MainClaim} (1).
\item{(III)}   After some steps, $\{\tilde{c}_{j}(n)\geq0\}$ are 
       sufficiently small for $j\neq 0$ and we can find  $t=t_{n}$ 
       such that
       $\alpha^{(n)}(t)=\alpha^{+(n)}(t)$ by using the implicit
       function theorem. This leads us to Claim \ref{MainClaim} (2).
\end{description}

\subsection{Problematic parts in the proof}

The previous proof may look perfect, but there must be 
something missing since this proof seems to work even for 
$G=U(1)$. We mention some problematic points:\\

\begin{description}
\item{Problem 1.}  The existence of $t_{m}$ satisfying 
(5.9) in \cite{tomb} or in (II) is not established.
\item{Problem 2.} The existence of $t_*$  satisfying 
\begin{equation}
\alpha_{\Lambda^{(n)},h}^{+(n)}(t_*)=\alpha_{\Lambda^{(n)},h}^{(n)}(t_*)
  \label{DREAM}
\end{equation}
is not established.
\item{Problem 3.} The implicit function theorem is used to obtain 
     $t(\lambda)$ which satisfies $\Psi(\lambda,t)=0$.
     It is not proven that $t(\lambda)$ 
     can be extended from $t=0$ to  $t=1$ even if $\Psi_{t}\neq 0$.
\end{description}

First we remark that all these claims depend on the hypothesis 
that both $\partial\alpha^{(n)}(t)/\partial t $ and 
$\partial \alpha^{+(n)}(t)/\partial t $ are bounded from 
below by strictly positive constants $\eta_{1}>0$ and
$\eta_{2}>0$ respectively, uniformly in $t\in R$.
But this cannot be correct without assuming 
$\alpha(t) \leq 1-\delta$ ($\delta>0$) 
for all $t$  uniformly in $n$ or $\Lambda^{(n)}$ 
because of the form  of $h(\alpha,t)=\exp[-t(1-\alpha)/\alpha]$. 
See (I). It is, however, very plausible that $\alpha$ is 
very close to $1$ since the original ($\alpha=1$) MK approximate
formulas  are very accurate. To ensure those lower bounds, the parameter 
$r$ is introduced. We suspect that this is the origin of the 
contradictions  as we explain in the next subsection.

Secondly  we need to find $t_{*}$ such that 
$\alpha(t_{*})=\alpha^{+}(t_{*})$, otherwise we compare two 
different physical systems, which is absolutely meaningless. 
To do so, the author of \cite{tomb} appeals to the 
implicit function theorem. But his claim does not follow from this 
theorem directly. We need that the map is contractive to
ensure that this claim is correct.

\subsection{The origin of the problems.}

The author of \cite{tomb} tries to ensure that 
$\partial \alpha/\partial t = v $ and 
$\partial \alpha^+/\partial t = v^+ $ are strictly bounded away 
from zero, and this is the reason for the introduction of the parameter 
$r$. The derivatives $v$ and $v^{+}$ contain
$1-\alpha$ in the numerator and $A_{\Lambda^{(m)}}$ in the denominator
which may make $v$ and $v^{+}$ small.
Here $A_{\Lambda^{(m)}}$ is (almost) equivalent to the derivative 
of the free energy by the (inverse) temperature
and is similar to the specific heat, see  (3.27) in Quotation (I) or 
in \cite{tomb}. It is easily seen that $A_{\Lambda^{(m)}}>0$ is bounded
from above  since $f(\{c_{j}\},U)\geq 0$. So the denominator is not 
dangerous.

Thus we have to worry about $1-\alpha\to 0$. 
 Since  the MK recursion formulas are very accurate for large 
 $\beta$, $\alpha$ seems to be very close to $1$ for large $\beta$. 
(This is so even though the MK recursion 
formulas fail to describe the deconfining transition of the $U(1)$ model 
even for $\alpha=r=1$.)  For this reason, the parameter $r\in(0,1]$ 
is introduced  by eq.(2.19) in \cite{tomb} so that 
$\alpha<1-\delta$. But the price to pay for this is that
$c_{j}(n)$ may not converge to 0 (the strong coupling fixed point) as  
$n\to\infty$ ! 

As remarked above, $c_{j}(n)=\hat{c}_{j}(n)^{b^{2}r}$ 
means that we replace  $b^{2}$ convolutions 
in (\ref{Convolution})  by $b^{2}r$ convolutions. 
If $r$ is chosen less than 1, 
the effective dimension of the MK recursion formula 
increases from $D=4$ to $D \geq 4$:
$$
D=4 \to D=4-\frac{2\log r}{2\log b+\log r}
$$
see \cite{ito, mk}. The choice $r=1$ corresponds to the original
MK  formula  and $D=4$ becomes the critical dimension
in the sense that the number of convolutions (i.e. $b^{2}$) is equal 
to the number of block plaquettes to be glued together (i.e. $b^{2}$).

If $r<1$, the recursion formula 
drives the system to the weak coupling fixed point if 
$\beta$ is large and $\alpha$ is close to 1.  Namely  if $r<1$ and 
$\beta$ is large, Theorem \ref{ito} does not hold and 
$\lim_{n\to\infty}c_{j}(n)\neq 0$ in general. 

The following pedagogical example makes this clear: Set 
$f_{0}(U)=\exp[-\beta\theta^{2}/2]$, 
where $\beta>0$ and $\theta\in R$. This corresponds to (non-compact)
abelian lattice QED. Then for the choice of $r\in (0,1]$, 
from (\ref{Convolution}), we get
$$
f_{0}\to f_{n}(\theta)=\exp\left[-\frac{\beta}{2r^{n}}\theta^{2}\right ]
$$   
which for $r<1$ converges to the weak coupling fixed point 
($\delta(\theta)$ with suitable normalization). 
If $r<1$, the same result will hold for compact $U(N)$ and $SU(N)$ 
lattice gauge theories for large $\beta$. This is plausible because in 
this regime the Gaussian approximation is good; it has been confirmed 
also numerically by us and \cite{suzuki}. As an illustration we 
show in the figure the results of numerical iteration for $G=SU(2)$ 
$r=0.9$ and $\alpha=1$. There is a `critical point' at 
$\beta\approx 4.8$: for $\beta\ge 4.8$ the modified MK flow goes to the 
weak coupling fixed point, whereas for $\beta\le 4.79$ it still flows to 
the strong coupling fixed point.

\begin{figure}[htb]
\includegraphics[width=.5\columnwidth]{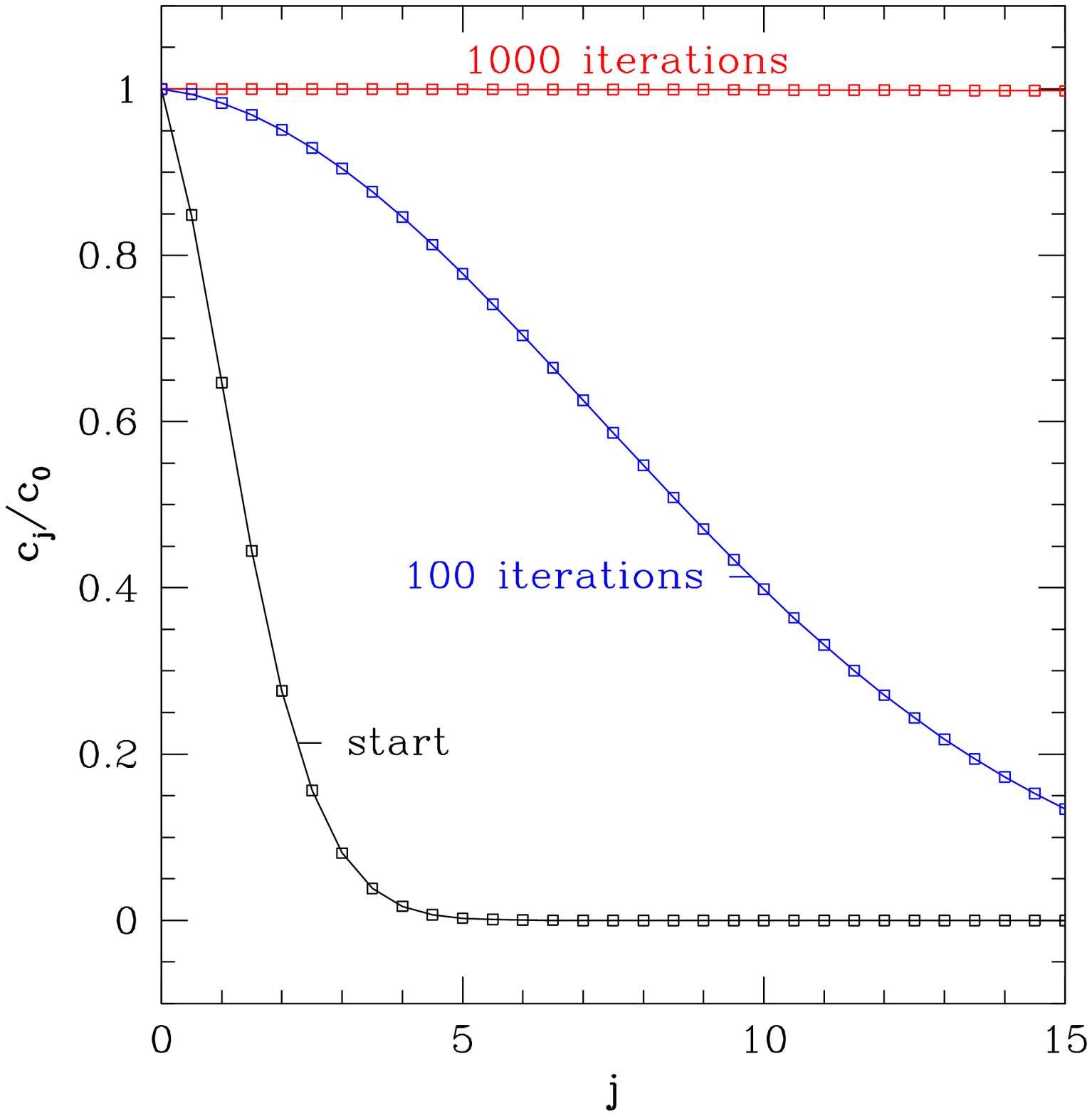}
\includegraphics[width=.5\columnwidth]{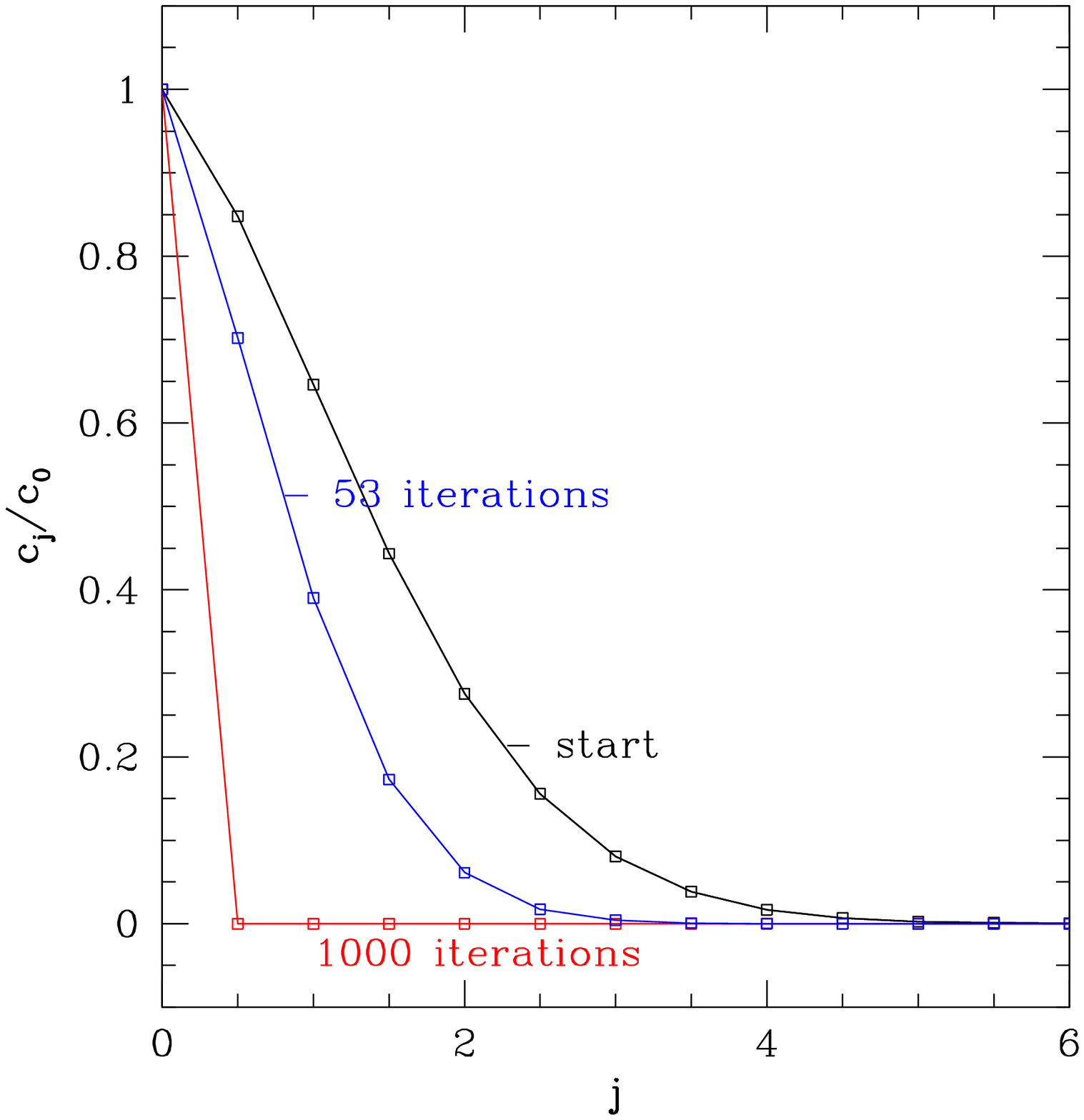}
\caption{Evolution of $c_j/c_0$ under Tomboulis' modified MK RG with 
$r=0.9$. $\beta=4.80$ (left plot), $\beta=4.79$ (right plot); lines drawn 
to guide the eye.
}
\label{plot}
\end{figure}

It should not be difficult to 
prove this rigorously. These examples show that one has to be very 
careful if $r$ is chosen $<1$: depending on $\beta$, $r$ has to be chosen
close to 1, which in turn will mean that $\alpha$ is close to 1. 

The author of \cite{tomb} says that ``choose $r<1$ if necessary 
$\ldots$ '' in several places (e.g. 3rd, 16th lines on  page 15, 
8th line on page 18, 15th line on page 23, etc.)
Though it is said that ``$r<1$ is actually irrelevant for $G=SU(2)$'' in
the 9th line on page 19, $r$ is chosen $<1$ throughout the paper. On the 
other hand, to  ensure $\lim_{n\to\infty} c_{j}(n)=0$, $1-r$ will have to 
be chosen very small for large $\beta$. But then the required upper bound 
$\alpha(r,t)<1-\delta$ becomes very subtle.

So we have two choices:
\begin{enumerate}
\item $r<1$. The larger $\beta$, the closer to one $r$ will have to 
       be chosen to make sure that the flow goes to the strong coupling 
       fixed point. But then it will be very difficult to ensure that
       $\alpha$ is sufficiently far from 1.	    	  
\item  $r=1$. All parameters depend on $\beta$ and $|\Lambda|$ in a 
       very subtle way in this case. (But this is sure to fail for
       $G=U(N)$.) 
\end{enumerate}
These subtle points are not addressed in \cite{tomb}.

Finally there is a problem with proving the existence of a $t$ 
such that $\alpha(t)=\alpha^{+}(t)$ after sufficiently many iterations of 
the MK recursion formulas or when $\{c_{j}\}$ are small. 
To obtain the function $t(\lambda)$ satisfying
$\Psi(\lambda,t)=0$ with $t(0)=0$, the author of \cite{tomb} appeals 
to the implicit function theorem.
Since 
\begin{eqnarray*}
\log Z^{+}(\{c_{j}\})
   &=& \log\left[F_{0}(1)^{h(\alpha^{+}(t),t)|\Lambda^{(1)}|}
                 Z_{1}^{+}(\{\tilde{c}(\alpha^{+}(t))\})\right]\\
   &=& \log\left[F_{0}(1)^{h(\alpha^{+}(t_{I}),t)|\Lambda^{(1)}|}
                 Z_{1}^{+}(\{\tilde{c}(\alpha^{+}(t_{I}))\})\right]
\end{eqnarray*}
by the parametrization invariance ( $t$-invariance) of the 
partition function (this is the definition of $\alpha^{+}$),
we have 
\begin{subequations}
\begin{eqnarray}
&& \Psi(\lambda=0,t) = h(\alpha(t),t)-h(\alpha^{+}(t_{I}),t_{I})\\
&& \Psi(\lambda=1,t)= 
         \frac{1}{\log F_{0}(1) |\Lambda^{(1)}|} \nonumber \\
&&\times 
         \left(\log \left[F_{0}(1)^{h(\alpha(t),t)|\Lambda^{(1)}|}
                 Z_{1}^{+}(\{\tilde{c}_{j}(\alpha(t))\})\right ]
                 -\log Z^{+}(\{c_{j}\}) \right )
\end{eqnarray}
\end{subequations}
We assume that the equation $\Psi(\lambda=0, t)=0$ is solved 
by $t=t_{0}$, and the equation $\Psi(\lambda,t)=0$ is our required 
equation, and we want to know if the solution $t=t(\lambda)$ with 
$t(0)=t_{0}$ can be continued to $t(1)$.  We have
\begin{eqnarray}
t(\lambda)&=& \phi  (t)(\lambda)
            \equiv  t_{0}+ \int_{0}^{\lambda}
                F(s,t(s)) ds \\
F(s,t(s))   &=& - \frac{\Psi_{s}(s,t(s))}{\Psi_{t}(s,t(s))}
\end{eqnarray}
where
\begin{subequations}
\begin{eqnarray}
\Psi_{t}(\lambda,t) &=&\left[1-
     \frac{h_{\alpha}(\alpha,t)+\lambda A^{+}(\alpha)}
          {h_{\alpha}(\alpha,t)+A(\alpha)} \right]
                     h_{t}(\alpha,t) \\
\Psi_{\lambda}(\lambda,t)
       & = &   \frac{1}{|\Lambda^{(1)}| \log F_{0}(1)} \nonumber \\
       & & \times
           \left(\log  Z_{1}^{+}(\{\tilde{c}_{j}(\alpha(t))\})
             -\log Z^{+}(\{\tilde{c}_{j}(\alpha^{+}(t)\}) \right )
\end{eqnarray}
\end{subequations}
and 
\begin{equation}
   h_{t}(\alpha,t)=-\frac{1-\alpha}{\alpha} h(\alpha,t), \quad 
   h_{\alpha}(\alpha,t)=\frac{t}{\alpha^{2}} h(\alpha,t)
\end{equation}
The integral equation $t(\lambda)=\phi(t)(\lambda)$ can be solved
analytically  by the iterations if $F(s,t)$ is bounded in the region. 
As Tomboulis pointed out in \cite{tomb}, $A\geq A^{+}$ if 
$\{c_{j}\geq 0\}$ are small and 
the high-temperature expansion converges. 
But the condition $\Psi_{t}(\lambda,t)\neq 0$ does not guarantee 
that $t(\lambda)$ can  be defined for all $\lambda$. The following 
example is given by Kanazawa \cite{Kana}:
$$
\Psi(\lambda,t)=e^{-t}-1+2\lambda
$$
To justify the claim in \cite{tomb}, we need contractivity of the 
map $\phi$ (in a suitable set of continuous functions) 
and to prove it is again not trivial at all.

\section {Discussion}
To sum up, the paper \cite{tomb} contains several problematic points 
which remain to be proved or confirmed. It seems that the remaining 
problems are, however, not easy to solve.

If the conventional wisdom of quark confinement in 4D non-abelian lattice 
gauge theory is correct, the alleged theorem in \cite{tomb} is certainly 
very plausible and may hold for $G=SU(N)$. But it is again a very subtle 
problem to show the existence of $t_*$ such that $\alpha(t_*)=\alpha^{+}(t_*)$
for a value of $r$ close enough to 1 to ensure convergence of the MK RG to 
the strong coupling fixed point.

The case of $G=U(1)$ highlights this subtlety: For large $\beta$ and $r=1$ 
such a $t_*$ does not exist, whereas for $r<1$ $t_*$ may exist, but 
does not imply confinement. We think in this note we made clear why it is 
so subtle to establish the existence of $t_*$.

Though the MK RG recursion formulae cannot distinguish non-abelian groups 
from abelian ones, the velocities of the convergences of 
$\{c_{j}(n)\}_{j=1/2}^{\infty}$ to 0 as $n\to\infty$ are very different. 
We are skeptical about the idea that the problem of quark confinement can 
be solved by soft analysis like this, but {\it if} the MK RG formulas 
should play a role in a rigorous proof of quark confinement in lattice 
gauge theory, these different velocities would certainly have to come into 
play. Presumably the dependence of $r$ and hence $\alpha$ on $\beta$ and 
$\Lambda$ must be clarified, but unfortunately it is very difficult to 
find out what the relationship is.

{\it {\bf Acknowledgments}. }
We would like to thank Prof. J.Greensite for useful correspondences
concerning Appendix C of \cite{tomb}. K.R.Ito would like to thank 
Dr.T.Kanazawa of Tokyo University who showed him the example 
for the implicit function theorem, and Prof.T.Yoneya and Prof.H.Suzuki
for useful discussions and computer simulation data on this problem.
K.R.I.'s work is financially supported by the Grant-in-Aid for 
Scientific Research (C) 17540208 from JSPS.


\end{document}